\documentclass[twocolumn,prb,superscriptaddress]{revtex4}

\usepackage{graphicx}
\usepackage{dcolumn}
\usepackage{bm}
\usepackage{amssymb}
\usepackage{amsmath}
\usepackage{color}

\begin{document}

\title{Anti-Kibble-Zurek behavior of a noisy transverse-field XY chain and its quantum simulation with two-level systems}

\author{Zhi-Peng Gao}

\affiliation{Guangdong Provincial Key Laboratory of Quantum Engineering and Quantum Materials,
SPTE, South China Normal University, Guangzhou 510006, China}

\author{Dan-Wei Zhang}
\email{zdanwei@126.com}\affiliation{Guangdong Provincial Key Laboratory of Quantum Engineering and Quantum Materials,
SPTE, South China Normal University, Guangzhou 510006, China}

\author{Yang Yu}
\affiliation{National Laboratory of Solid
State Microstructures and School of Physics, Nanjing University,
Nanjing 210093, China}

\author{Shi-Liang Zhu}
\email{slzhu@nju.edu.cn}
\affiliation{National Laboratory of Solid
State Microstructures and School of Physics, Nanjing University,
Nanjing 210093, China}
\affiliation{Guangdong Provincial Key Laboratory of Quantum Engineering and Quantum Materials,
SPTE, South China Normal University, Guangzhou 510006, China}
\affiliation{Synergetic Innovation Center
of Quantum Information and Quantum Physics, University of Science
and Technology of China, Hefei, Anhui 230026, China}

\begin{abstract}
We study the dynamics of a transverse-field XY chain driven across
quantum critical points by noisy control fields.  We characterize
the defect density as a function of the quench time and the noise
strength, and demonstrate that the defect productions for three
quench protocols with different scaling exponents exhibit the
anti-Kibble-Zurek behavior, whereby slower driving results in more
defects. The protocols are quenching through the boundary line
between paramagnetic and ferromagnetic phases, quenching across
the isolated multicritical point and along the gapless line,
respectively. We also show that the optimal quench time to
minimize defects scales as a universal power law of the noise
strength in all the three cases. Furthermore, by using quantum
simulation of the quench dynamics in the spin system with
well-designed Landau-Zener crossings in pseudo-momentum space, we
propose an experimentally feasible scheme to test the predicted
anti-Kibble-Zurek behavior of this noisy transverse-field XY chain
with two-level systems under controllable fluctuations.
\end{abstract}

\date{\today}

\maketitle

\section{introduction}

Kibble-Zurek mechanism (KZM) provides an elegant theoretical
framework for exploring the critical dynamics of phase transitions
in systems ranging from cosmology to condensed matter
\cite{Kibble1976,Zurek1985,Zurek1996}.  The dynamics induced by a
quench across a critical point with a control parameter is
generally nonadiabatic due to the critical slowing down, which
results in the production of topological defects. A key prediction
of KZM is that the density of defects $n_0$ follows a universal
power law as a function of the quench time $\tau$ (transition rate
$1/\tau$): $n_0\propto \tau^{-\beta}$, where the scaling exponent
$\beta>0$ determined by the critical exponents of the phase
transition and the dimensionality of the system. KZM for classical
continuous phase transitions has been verified in many systems,
such as cold atomic gases \cite{Navon2015}, ion crystals
\cite{Ulm2013,Pyka2013}, and superconductors \cite{Monaco2002}.
There has been significant theoretical work on extension of KZM
for quantum phase transitions
\cite{Dziarmaga2010,Isingexact1,Isingexact2,Dziarmaga2006,Caneva2007,Xy1,XyMC,Xy2,Deng2009,Sabbatini,Kolodrubetz,Caneva2008,Acevedo2014},
which are zero temperature transitions driven by Heisenberg
quantum fluctuations rather than thermal fluctuations \cite{QPT}.
For instance, by studying KZM in the one-dimensional
transverse-field Ising model, which is one of the paradigmatic
models to study quantum phase transitions, it was found that the
density of defects scales as the square root of the quench time
with the scaling exponent $\beta=1/2$
\cite{Dziarmaga2010,Isingexact1,Isingexact2}. However, the
experimental tests of KZM in quantum phase transitions are still
scare since controlling the time evolution of systems cross
quantum critical points is notoriously difficult
\cite{Chen2011,Braun,Anquez,Chin}.

Landau-Zener transition (LZT), occurring when a two-level system
sweeps through its anticrossing point, has served over decades  as
a textbook paradigm of quantum dynamics of some non-equilibrium
physics \cite{Landau,Zener}. Recently, LZT has been extensively
studied \cite{Shevchenko} both theoretically and experimentally
in, e.g., superconducting qubits \cite{Oliver,Sillanpaa,Tan},
solid-state spin systems \cite{Petta,Betthausen,Cao}, and optical
lattices \cite{Tarruell,Salger,Chen}. It was shown that the
dynamics of LZT can be intuitively described in terms of KZM of
the topological defect formation in non-equilibrium quantum phase
transition \cite{LZKZM1,LZKZM2}. The correspondence between the
two physical situations provide a promising way for
proof-of-concept quantum simulation of KZM in quantum regime by
using LZT in two-level systems, which has been experimentally
demonstrated in an optical interferometer \cite{Xu}. Moreover,
quantum simulation of the critical dynamics in the
transverse-field Ising model by a set of independent Landau-Zener
crossings in pseudo-momentum space has been realized in a
semiconductor electron charge qubit \cite{LZKZMexp_QD}, a
superconducting qubit  \cite{LZKZMexp_SC} and a single trapped ion
\cite{LZKZMexp_Ion}. The LZT there can be engineered well and
probed with high accuracy and thus KZM of defect production in the
Ising model with the scaling exponent $\beta=1/2$ has been
successfully observed in these artificial two-level systems
\cite{LZKZMexp_QD,LZKZMexp_SC,LZKZMexp_Ion}.

While KZM has been verified to be broadly applicable, a
conflicting observation was reported in a recent experiment of
ferroelectric  phase transition: slower quenches generate more
defects when approaching the adiabatic limit \cite{Griffin}. This
behavior is opposite to that predicted by the standard KZM and is
termed as anti-Kibble-Zurek (anti-KZ) behavior. The quench
dynamics of a transverse-field Ising chain coupled to a
dissipative thermal bath has been theoretically studied in Refs.
\cite{Santoro,Nalbach}, which show that the defect production can
exhibit the anti-KZ behavior due to the emergence of thermal
defects. Recently, by studying the crossing of the quantum
critical point in a thermally isolated Ising chain driven by a
noisy transverse field, it was demonstrated that noise
contributions can also give rise to anti-KZ behavior when they
dominate the dynamics \cite{Dutta}. A natural question is whether
the anti-KZ behavior can exhibit in other quantum spin models with
different scaling exponents under noisy control fields. In the
experimental aspect, it would be of great value to set a stage for
quantum simulation of such anti-KZ behavior in two-level systems
with Landau-Zener crossings, noting that quantum spin models can
only be realized in some special situations without controllable
noisy fields \cite{Kim,Bohnet,Simon}, which may prevent the test
of the anti-KZ behavior.

In this paper, we consider the dynamics of a transverse-field XY
chain driven across quantum critical points by noisy control
fields.  We numerically calculate the defect density as a function
of the quench time and the noise strength, and find that the
defect productions in  three quench protocols with different
scaling exponents exhibit the anti-KZ behavior, i.e., slower
driving results in more defects. The three protocols are quenching
through the boundary line between paramagnetic and ferromagnetic
phase, quenching across the isolated multicritical point and along
the gapless line, which have the Kibble-Zurek scaling exponents
$\beta=1/2,1/6,1/3$ under the noise-free
driving \cite{Xy1,XyMC,Xy2}, respectively. We also show that the
optimal quench time to minimize defects scales as a universal
power law of the noise strength in all three cases. Furthermore,
by using quantum simulation of the quench dynamics in the spin
system with well-designed Landau-Zener crossings, we then propose
an experimentally feasible scheme to test the predicted anti-KZ
behavior in this noisy transverse-field XY chain with two-level
systems under controllable fluctuations in the control field. The
driving protocols of the Landau-Zener crossings in two-level
systems and the required parameter regions for observing the
anti-KZ behavior in the three cases are presented.

The paper is organized as follows. In Section II, we illustrate
that the defect productions of the  transverse-field XY chain
driven across quantum critical points by noisy control fields
exhibit the anti-KZ behavior. In Section III, we propose to test
the predicted anti-KZ behavior by using quantum simulation of the
quench dynamics with well-designed LZT in two-level systems.
Finally, a short conclusion is given in Sec. IV.

\section{Anti-KZ behavior of a noisy transverse-field XY chain}

We begin with the spin-$1/2$ quantum  XY chain under a uniform
transverse field (homogeneous for each spin), which is one of the other exactly solvable spin
models apart from the quantum Ising chain. The Hamiltonian of the
transverse-field XY chain with nearest neighbor interaction is
given by \cite{Lieb1961,Bunder1999}
\begin{eqnarray}\label{Hxy1}
H=-\frac{1}{2}\sum_{n=1}^{N}\left(J_{x}\sigma_{n}^{x}\sigma_{n+1}^{x}+J_{y}\sigma_{n}^{y}\sigma_{n+1}^{y}+h\sigma_{n}^{z}\right),
\end{eqnarray}
where $N$ (here and hereafter we set $N$ is even) counts the
number of spins, $\sigma_{n}^{i}$ ($i=x,y,z$) are the Pauli
matrices acting on the $n$-th spin, $J_{x}$ and $J_{y}$
respectively represent the anisotropy interactions along $x$ and
$y$ spin directions, $h$ measures the strength of the transverse
field. We set $J=J_{x}+J_{y}$, $\gamma=(J_{x}-J_{y})/J$, then the
Hamiltonian can be rewritten as
\begin{eqnarray}
\nonumber H&=&-\frac{J}{2}\sum_{n=1}^{N}\left[(1+\gamma)\sigma_{n}^{x}\sigma_{n+1}^{x}+(1-\gamma)\sigma_{n}^{y}\sigma_{n+1}^{y}\right] \\
&&-h\sum_{n=1}^{N}\sigma_{n}^{z}. \label{Hxy2}
\end{eqnarray}
The system reduces to the isotropic XY chain for $\gamma=0$ and
the Ising chain for $\gamma=1$. This Hamiltonian can be exactly
diagonalized by using the Jordan-Wigner transformation, which maps
a system of spin-1/2 to a system of spinless free fermions
\cite{Lieb1961,Bunder1999,Caneva2007}. The Jordan-Wigner
transformation of spins to fermions is given by $\sigma_{n}^{\pm}
= \exp\left(\pm i\pi\sum_{m=1}^{n-1}c_{m}^{\dag}c_{m}\right)c_{n}$
and $\sigma_{n}^{z} = 2c_{n}^{\dag}c_{n}-1$, where
$\sigma_{n}^{\pm}=\sigma_{x}\pm i\sigma_{y}$. In the fermionic
language, the XY model Hamiltonian can be rewritten as
\begin{eqnarray}\label{HJW}
\nonumber H & = & -J\sum_{l=1}^{N}\left[(c_{l}^{\dag}c_{l+1}+c_{l+1}^{\dag}c_{l})+\gamma(c_{l}^{\dag}c_{l+1}^{\dag}+c_{l+1}c_{l})\right]\\
 &  & -h\sum_{l=1}^{N}\left(2c_{l}^{\dag}c_{l}-1\right).
\end{eqnarray}
Under the periodic boundary condition
$\sigma_{N+1}^{\alpha}=\sigma_{1}^{\alpha}$ with even $N$ requires
that $c_{N+1}=-c_{1}$ and after the Fourier transformation with
$c_{n}=\frac{e^{-i\pi/4}}{\sqrt{N}}\sum_{k\in(-\pi,\pi]}\left(e^{ikn}c_{k}\right)$,
one can obtain $H=\sum_{k\in[0,\pi]}\Psi_{k}^{\dag}\mathcal{H}(k)
\Psi_{k}$, where $\Psi_{k}^{\dag}=(c_{k}^{\dag},c_{-k})$ and the
Hamiltonian density in the pseodu-momentum space
\begin{equation}
\mathcal{H}(k)=-2\left[\hat{\sigma}_{z}(J\cos k+h)+\hat{\sigma}_{x}(J\gamma\sin k)\right].\label{HFT}
\end{equation}
Note that here and hereafter the Pauli matrices
$\hat{\sigma}_x,\hat{\sigma}_z$ are conventionally used to write
the $2\times2$ Hamiltonian of each independent $k$-mode ($\{-k,k\}$ pair), which is
distinct from the spin components $\sigma_n^i$ in Eqs.
(\ref{Hxy1}) and (\ref{Hxy2}).

\begin{figure}[tbph]
\centering
\includegraphics[width=7cm]{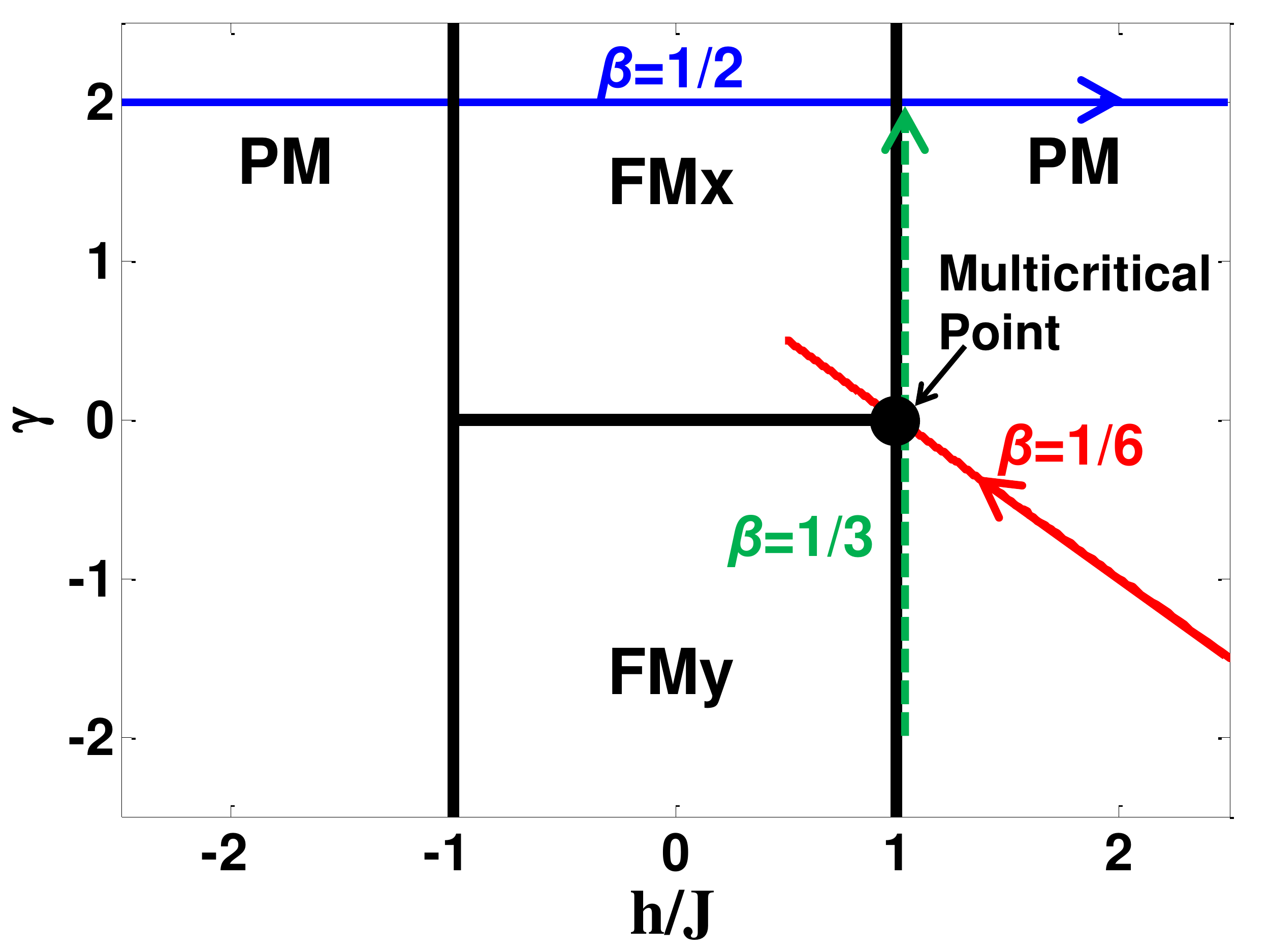}
\caption{(Color online) Phase diagram of the transverse-field XY
chain and three quench protocols. The two vertical black solid
lines separate the $\gamma-h/J$ plane into three parts. For
$|h|>|J|$, the XY chain is in the paramagnetic phase denoted by
PM. The middle area $|h|<|J|$ is divided into two parts (separated
by horizontal black solid line at the center), which are two
ferromagnetic phases denoted by FMx and FMy, ordering along $x$
and $y$ directions for $J_x>J_y$ and $J_x<J_y$, respectively. The
horizontal blue line represents the path of the transverse quench
with Kibble-Zurek scaling exponent $\beta=1/2$. The anisotropic
quench through the multicritical point (the red point) with
$\beta=1/6$ is denoted by red line across the multicritical point,
and the third quench along the gapless line ($h=J$) with
$\beta=1/3$ is denoted by the green dotted line.}
\label{phase-diagram}
\end{figure}

The energy spectrum of the system can be obtained by
diagonalizing Eq. (\ref{HFT}), and thus the critical points or
lines between different quantum phases can be found by minimizing
the energy gap. Figure \ref{phase-diagram} depicts the phase
diagram of the transverse-field XY chain in the space spanned by
the parameters $h/J$ and $\gamma$ \cite{Xy1}. There are a quantum
paramagnetic phase denoted by PM and two ferromagnetic long-ranged
phases ordering along $x$ and $y$ directions denoted by FMx and
FMy, respectively. As shown in Fig. \ref{phase-diagram}, there are
three kinds of phase phase boundaries: an Ising critical line on
the horizontal axis $\gamma=0$ (i.e. $J_x=J_y$), a multicritical
point locates at $\gamma=0, h/J=1$, and two gapless lines along
$h/J=\pm 1$ \cite{Xy1,XyMC,Xy2}. Quantum phase transition occurs
when the system is driven by one or more parameters to across the
boundary line or point on the phase diagram.

\begin{figure*}[tbph]
\centering
\includegraphics[width=17.8cm,height=9.2cm]{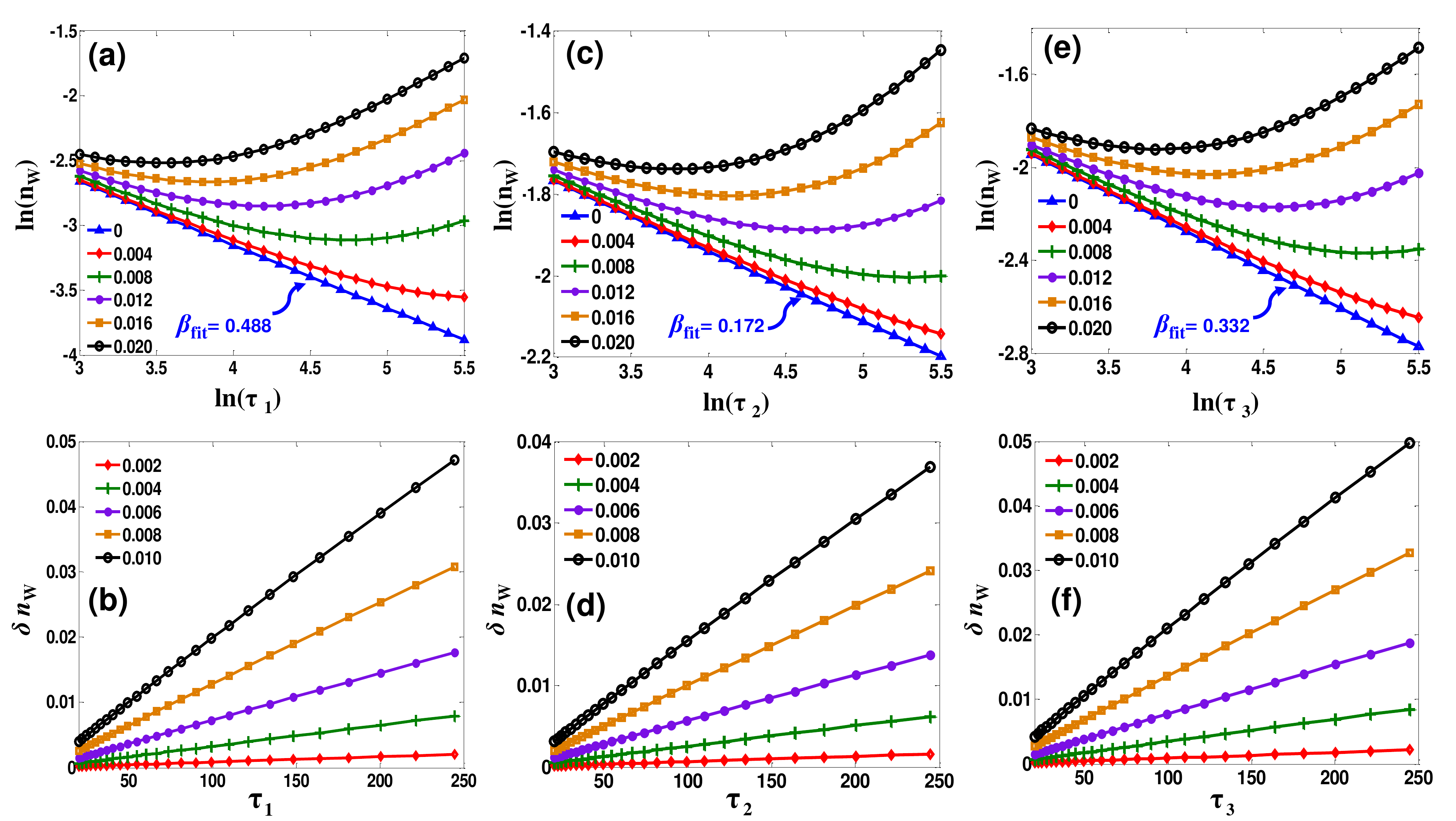}
\caption{(Color online) The anti-KZ behavior of the defect
productions in three quench protocols. (a,c,e) For the three
protocols $j=1,2,3$, respectively, the defect density
$n_w(\tau_j)$ as a function of the quench time for different noise
strength $W$. (b,d,f) The corresponding noise-induced defect
density $\delta n=n_w-n_0$ in the three cases. The quantity of $W$
is marked by legend. The three quench protocols are the transverse
quench ($j=1$) with the linear fitting scaling exponent
$\beta_{\text{fit}}=0.488$ in the noise-free limit $W=0$, the
anisotropic quench across the multicritical point ($j=2$) with
$\beta_{\text{fit}}=0.172$, and the quench along the gapless line
($j=3$) with $\beta_{\text{fit}}=0.332$. The larger $W$ causes
larger divergence from the power law prediction in KZM and the
systems exhibit the anti-KZ behavior, whereby slower driving
results in more defects and is remarkable in the long quench time
limit. $N_k=500$ is set in the simulations.} \label{Anti-KZ}
\end{figure*}

To study the quench dynamics of the transverse-field XY chain, we
can use the Hamiltonian  decoupling into a sum of independent
terms $H(t)=\sum_{k\in[0,\pi]}\mathcal{H}(k,t)$, where each
$k$-mode Hamiltonian $\mathcal{H}(k,t)$ given by Eq. (\ref{HFT})
operates on a two-dimensional Hilbert space. The time evolution of
a generic state is governed by the Schr\"{o}dinger equation
$i\frac{d}{dt}|{\psi_k(t)\rangle}=\mathcal{H}(k,t)|\psi_k(t)\rangle$.
This projection of the spin Hamiltonian to the $2\times2$ Hilbert
space has effectively reduced the quantum many-body problem to the
problem of an array of decoupled two-level systems.

For the convenience of experimental implementation, we choose only
one of the parameters $h,J_x,J_y$ and $\gamma$ as  linearly
quenched in time and the rest fixed in a specific quench protocol.
As shown in Fig. \ref{phase-diagram}, we consider three different
quench protocols for the system driven across the phase
boundaries: (1) The transverse quench with only $h(t)$ is varied
in time, which is quenching through the boundary line between
paramagnetic and ferromagnetic phase twice; (2) The anisotropic
quench across the isolated multicritical point, in which case only
the parameter $J_x(t)$ is varying and $h=2J_y$ is set to ensure
passing through the multicritical point; (3) The quench along the
gapless line, which requires that $h/J=1$ and only $\gamma(t)$ is
varied. For the linear quench in the absence of dissipative
thermal bath or noise fluctuations, the density of defects [see
Eq. (\ref{defect})] formed in three quench protocols follows the
Kibble-Zurek power law as a function of the quench time with the
scaling exponents $\beta=1/2,1/6,1/3$ \cite{Xy1,XyMC,Xy2},
respectively. In the following, we consider the three quench
protocols in the transverse-field XY chain under noisy control
fields and demonstrate the exhibition of anti-KZ behavior.

We now present a general framework for the description of the
noise fluctuations denoted by $\eta(t)$ in the control fields.
The total quench parameter is written as
\begin{eqnarray}
f_j(t) &=& f^{(0)}_j(t)+\eta(t),
\end{eqnarray}
where $f^{(0)}_j(t)\propto t/\tau_j$ denote perfect control
parameter linearly varying in time with quench time $\tau_j$ and
the subscript $j=1,2,3$ respectively represent the three quench
protocols. Here $\eta(t)$ is white Gaussian noise with zero mean
and the second moment
$\langle{\eta(t)\eta(t')}\rangle\rangle=W^{2}\delta(t-t')$, with $W^2$ being
the strength of noise fluctuation (here $\eta$ is
dimensionless and $W^2$ has units of time). Note that white noise
is a good approximation to ubiquitous colored noise with
exponentially decaying correlations. We set $W$ as a small value for the stochastic perturbation,
which ensures the validity of noise-average density matrix
technique \cite{Armin2015}. We consider the system Hamiltonian
containing two parts in a general form \cite{Dutta}
\begin{equation}\label{Ham_two_part}
H(t)=H^{(0)}(t)+\eta(t)V,
\end{equation}
where $H^{(0)}(t)$ denotes the ideal quench Hamiltonian to
describe the prescheduled evolution of the driven system, and $\eta(t)V$
denotes the fluctuation of the control fields which
modify the driving process as an effectively open quantum
dynamics. Defining the stochastic wave
function $|\psi_{\eta}(t)\rangle$, the stochastic Schr\"{o}dinger equation
is applied to describe the interplay of these two factors in one noise realization (let $\hbar=1$):
\begin{equation}\label{SSE}
i\frac{d}{dt}|{\psi_{\eta}(t)\rangle}=[H^{(0)}(t)+\eta(t)V]|\psi_{\eta}(t)\rangle.
\end{equation}
The stochastic density matrix
$\rho_{\eta}=|\psi_{\eta}\rangle\langle\psi_{\eta}|$ is a function
of $\eta(t)$, whose equation of motion can be derived from the
dual pair of the stochastic Schr\"{o}dinger equation and is given
by
\begin{eqnarray}
\frac{d}{dt}\rho_{\eta}(t)=-i[H^{(0)}(t),\rho_{\eta}(t)]-i[V,\eta(t)\rho_{\eta}(t)].
\end{eqnarray}
We assume that all noise realizations are independent,
which is a practical situation in realistic experiments. This
allows us to implement an average over noise realizations in the
stochastic process. We denote the noise-averaged density matrix by
$\rho(t)=\langle{\rho_{\eta}(t)}\rangle$, which is a solution of
the master equation
\begin{eqnarray}
\frac{d}{dt}\rho(t)=-i[H^{(0)}(t),\rho_{\eta}(t)]-i[V,\langle\eta(t)\rho_{\eta}(t)\rangle].
\end{eqnarray}
By using Novikov's theorem \cite{Novikov} for the considered
Gaussian noises, one can find
$\langle\eta(t)\rho_{\eta}(t)\rangle=-iW^2[V,\rho(t)]/2$ and
obtain the nonperturbative exact master equation given by
\cite{Dutta}
\begin{eqnarray}\label{eq:rho_k}
\frac{d}{dt}\rho(t)=-i[H^{(0)}(t),\rho(t)]-\frac{W^2}{2}[V,[V,\rho(t)]].
\end{eqnarray}
This equation is directly related to the detection in realistic
experiments and thus can be used to perform simulations of
the three quench protocols under noisy control fields with the
strength parameter $W$.

The definition of the density of defects in the transverse field
XY chain after the quench is straightforward, similar to the case
for the Ising model \cite{Dziarmaga2010,Isingexact1,Isingexact2}.
For each $k$-mode in a specific quench protocol, one can
numerically simulate the time evolution and find the
noise-averaged density matrix $\rho_{k}(\tau_j)$ at the end of
quench (with the quench time $\tau_{j}$). In the basis of
adiabatic instantaneous eigenstate
$\{|G_{k}(\tau_j)\rangle,|E_{k}(\tau_j)\rangle\}$, one has
$p_{k}=\langle
E_{k}(\tau_j)|\rho_{k}(\tau_j)|E_{k}(\tau_j)\rangle$ to measure
the probability in the excited state $|E_{k}(\tau_j)\rangle$ of
each $k$-mode. For the whole XY chain, in other equivalence words,
all $k$-modes contribute to the density of defects $n_{W}$ (the
subscript $W$ denotes the presence of noises) as reasonably
defined by
\begin{equation} \label{defect}
n_{W}=\frac{1}{N_{k}}\sum_{k\in[0,\pi]}p_{k},
\end{equation}
where $N_{k}$ is the number of $k$-modes used in the summation,
and it is consistent of $n_0$ for noise-free case with $W=0$.

Utilizing the theoretical framework, three quench protocols can be
analyzed by substituting specific $\mathcal{H}_{j}^{(0)}(t)$ and
$V_{j}(t)$ into Eq. (\ref{eq:rho_k}). We first consider the
transverse quench ($j=1$), in which case only the parameter
$h(t)$ is time-dependent, as shown in Fig. \ref{phase-diagram}.
To describe the effect of noise in the control field $h$, the
Hamiltonian for each $k$-mode $\mathcal{H}_{1}(k)$ can be
separated into the determined and stochastic parts as
$\mathcal{H}_{1}(k)=\mathcal{H}_{1}^{(0)}(k)+V_1$, where
\begin{eqnarray}
\nonumber \mathcal{H}_{1}^{(0)}(k) & = & -2[(J_x+J_y)\cos k+h(t)]\hat{\sigma}_{z}\\
\nonumber  && -2[(J_x-J_y)\sin k]\hat{\sigma}_{x},\\ \label{H_h0}
V_1 & = & -2\hat{\sigma}_{z}.
\end{eqnarray}
Here $h(t)=f^{(0)}_1=v_ht$ with the quench velocity $v_h$ drives
the system from left PM-phase region through middle FM-phase part
to the right PM-phase region as shown in Fig. \ref{phase-diagram}.
In our numerical simulations, $h(t)$ is set to vary from
$h_{\text{min}}=-5/3$ to $h_{\text{max}}=5/3$, with the other two
independent parameters being fixed as $J_x=1$ and $J_y=-1/3$. For
the entire process of the quench time $\tau_1$, we have
$v_h=(h_{\text{max}}-h_{\text{min}})/\tau_{1}=10/(3\tau_1)$ and the whole
evolution time $t\in[-\tau_1/2,\tau_1/2]$.

Figure \ref{Anti-KZ}(a) illustrates the numerical results of the
defect density $n_W$ as a function of the quench time $\tau_1$ in
the transverse quench protocol. When the driving field is free
from noises with $W=0$, the results of
$n_0\propto\tau_1^{-\beta_{\text{fit}}}$ with the fitting exponent
$\beta_{\text{fit}}=0.488$ agrees well with the theoretical
prediction of Kibble-Zeruk scaling exponent $\beta=1/2$ in the
thermodynamical and long quench time limits \cite{Xy1,XyMC,Xy2}.
Note that the deviation of $\beta_{\text{fit}}$ (about
$2\%$) here comes from the finite size of $k$-modes $N_k$
($N_k=500$ is set) and the finite quench time $\tau_1$ in our
simulations. We have numerically confirmed that this deviation can
be decreased with the increasing of $N_k$ and $\tau_1$. For short
quench time, the defect density $n_W$ as a function of the quench
time $\tau_1$ is close to the scaling form in the noise-free case.
As the strength of noise fluctuation grows (increasing $W$), the
corresponding defect density increases compared to the results
under noise-free condition, and the deviation becomes more
significant for longer quench time, where the dynamics is
dominated by the noise-induced non-adiabatic effects. Then the
power-law scaling form of $n_W(\tau_1)$ fails and the system
exhibits the anti-KZ behavior, i.e., slower driving (larger
$\tau_1$) results in more defects. Finally $n_W$ is completely
governed by the anti-KZ contribution in the limit of very long
quench time.

The physics of the anti-KZ behavior can be further
interpreted as follows: The defect production of the noise-free
part of the quench system is due to the critical slowing down in
KZM. In contrast, the noisy fluctuations in the control fields
allow the absorption of energy to generate excitations in the
system, which accumulate during the evolution with the increasing
of the quench time. Thus, the resulting defect production is
determined by the two independent mechanisms. For the relatively
weak noises considered in our work, the noises contribute
negligible excitations for the short quench time and then the
scaling of the defect production is still effectively governed by
the Kibble-Zurek predictions. When the quench time becomes large
enough, the accumulation of noise-induced excitations dominates
and the system enters the anti-KZ regime.

\begin{figure}[tbph]
\includegraphics[width=7cm]{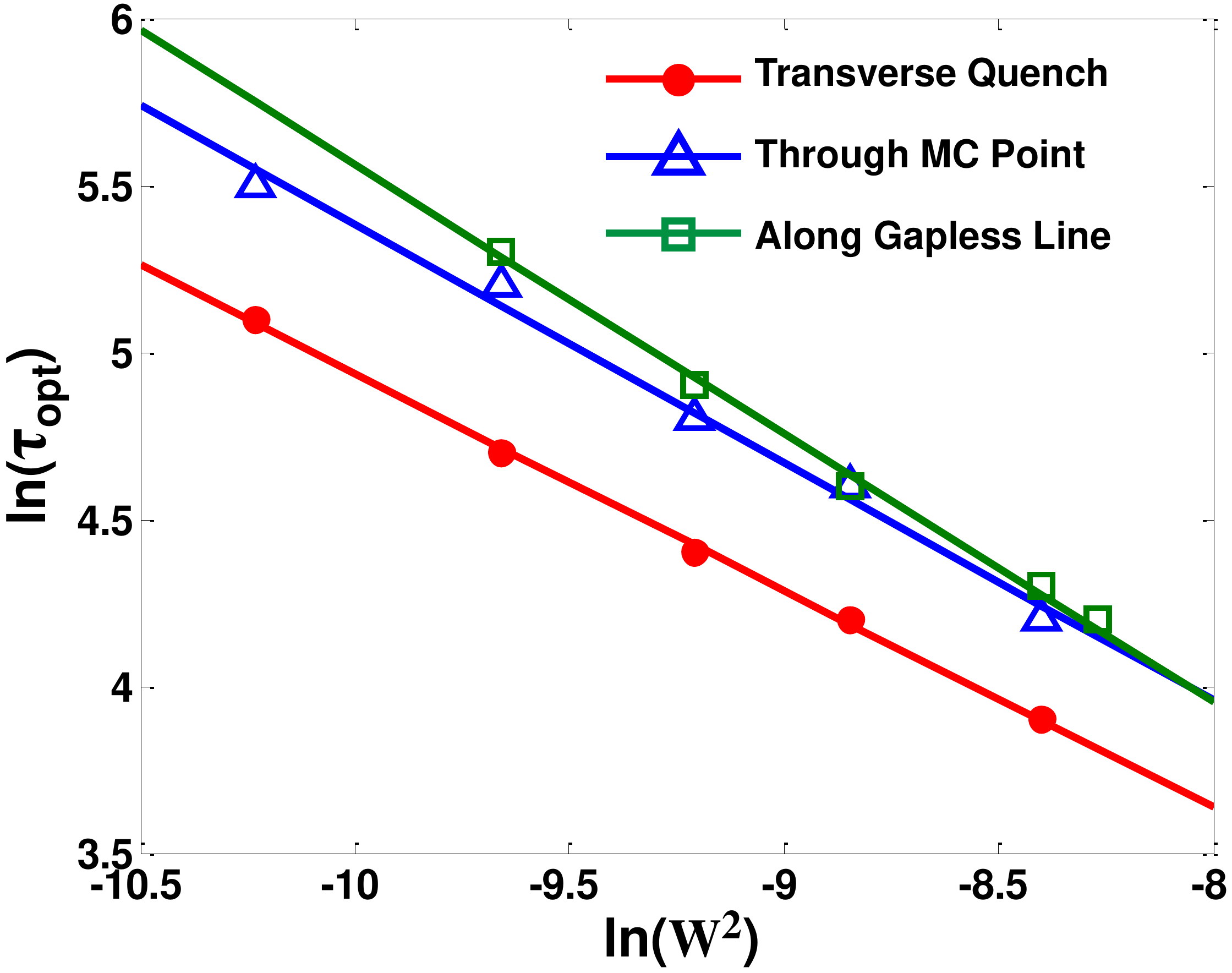}
\caption{(Color online) The optimal quench time
$\tau_{\text{opt}}$ to minimize defects scales as a power law of
the noise strength in the three quench protocols.
$\ln(\tau_{\text{opt}})$ as a function of $\ln(W^{2})$ with linear
fitting  gives $\ln(\tau_{\text{opt}})\propto
\alpha_{\text{fit}}\ln(W^{2})$, where the fitting parameters for
the three cases are given by: $\alpha_{\text{fit}}=-0.6496$ for
the transverse quench, which is close to analytical result
$\alpha=-2/3=-0.6667$; $\alpha_{\text{fit}}=-0.8055$ for the
quench through the multicritical point with $\alpha=-6/7=-0.8571$;
$\alpha_{\text{fit}}=-0.71196$ for the quench along gapless line
with $\alpha=-3/4=-0.75$.} \label{OptimalTime}
\end{figure}

We proceed to consider the second quench protocol ($j=2$), the
anisotropic quench across the multicritical point with the control
field $J_{x}(t)$ and fixed $h=2J_y$, as shown in Fig.
\ref{phase-diagram}. In this case, the Hamiltonian for each
$k$-mode can be written as
$\mathcal{H}_{2}(k)=\mathcal{H}_{2}^{(0)}(k)+V_2$, where
\begin{eqnarray}\label{HJx}
\nonumber \mathcal{H}_2^{(0)}(k)&=& -2\{[J_x(t)+J_y]\cos(k)+h\}\hat{\sigma}_z \\
\nonumber &&-2[(J_x(t)-J_y)\sin(k)]\hat{\sigma}_x,  \\
    V_2&=& -2[(\sin k)\hat{\sigma}_x+(\cos k)\hat{\sigma}_z].
\end{eqnarray}
Here the linearly driving field $J_x(t)=f^{(0)}_2(t)=v_xt$ with the
quench velocity $v_x$. Similar to the first protocol mentioned
above, in our numerical simulations, we choose $h=2$ and $J_y=1$,
and let $J_x$ ramp from $-1$ to $3$ with the overall quench time
$\tau_2$. Under this condition, the system is initially in the PM
phase and then driven through the multicritical point into the FMx
phase. Consequently, the quench velocity $v_x=4/\tau_2$, and the
evolution progress is $t\in[-\tau_2/4,3\tau_2/4]$. Figure
\ref{Anti-KZ}(c) shows the numerical results of the defect density
$n_W(\tau_2)$ in this quench protocol. In the noise-free limit
with $W=0$, we find that $n_0\propto\tau_1^{-\beta_{\text{fit}}}$
with the fitting exponent $\beta_{\text{fit}}=0.172$ agrees with the theoretical prediction
$\beta=1/6$ \cite{Xy1,XyMC,Xy2}. When the dynamics is dominated
by the noise effects for increasing the noise strength and (or)
quench time, this power-law scaling again fails and the system
exhibits the anti-KZ behavior.

For the last quench protocol along the gapless line protocol
($j=3$), the parameter $\gamma=f^{(0)}_3=(J_{x}-J_{y})/J$ is used
as the control field to linearly drive the system, which takes the
form $\gamma(t)=v_{\gamma}t=(4/\tau_3) t$ in the evolution
progress $t\in[-\tau_3/2,\tau_3/2]$. The other parameters are set
as $h=J=J_{x}+J_{y}=1$. Therefore, the Hamiltonian for each
$k$-mode in this case is
$\mathcal{H}_{3}(k)=\mathcal{H}_{3}^{(0)}(k)+V_3$, where the two
parts
\begin{eqnarray}\label{H3}
\nonumber \mathcal{H}_{3}^{(0)}(k) & = & -2[(J\cos k+h)\hat{\sigma}_{z}+J\gamma(t)\sin k\hat{\sigma}_{x}],\\
V_3 & = & -2J\sin k\hat{\sigma}_{x}.
\end{eqnarray}
The numerical results of the defect density $n_W(\tau_3)$ in this
quench protocol are shown in Fig. \ref{Anti-KZ}(e). For $W=0$, we
find that $n_0\propto\tau_1^{-\beta_{\text{fit}}}$ with the
fitting scaling exponent $\beta_{\text{fit}}=0.332$ consistent
with the theoretical prediction $\beta=1/3$
\cite{Xy1,XyMC,Xy2}. Increasing the noise strength $W$, the system
inters the anti-KZ regime, with more defects formed for longer
quench time. Hence, we come to the conclusion that when the noises
presents in the control fields, the anti-KZ behavior exists in all
the three quench protocols in the transverse-field XY chain with
different noise-free Kibble-Zeruk scaling exponents.

Due to the exhibition of the anti-KZ behavior in the quench when
the noise presents, it is imperative to find an optimal quench
time $\tau_{\text{opt}}$ to minimize the defects. Under the
condition of finite quench time $\tau_{j}$, the optimal control in
the annealing of quantum simulator give a challenge that defect
(excitation) density is produced as less as possible. For our
numerical results, we focus on the region $3.0\leq
\ln(\tau_{j})\leq5.5$, since the experimental systems only
maintain their quantum coherent characteristic in finite ramp
time. As the defect density in the absence of noises $n_{0}\approx
c\tau_{j}^{-\beta}$, where the prefactor $c$ is predicted by KZM
and depends on s specific protocol. One can argue
that in the limit of small noise and finite quench time
\cite{Dutta}, the total density of defect $n_{W}\approx
n_{0}+\delta n$, where the noised-induced part $\delta n$ is given
by
\begin{eqnarray}\label{delta_n}
\delta n & \approx & n_{W}-c\tau_{j}^{-\beta}.
\end{eqnarray}
Note that the effective decoupling of the KZM dynamics from
noise-induced effects, as interpreted previously, leads to the
additive form of $n_{W}$. Figure 2(b,d,f) display that $\delta
n$ for the three quench protocols ($j=1,2,3$) is almost linearly
depend on $\tau_j$ when $W\ll1$, and thus one has
\begin{eqnarray}
\delta n\approx r\tau_j,
\end{eqnarray}
with $r$ being the coefficient whose value depends on the noise strength.
Secondly, the optimal quench time $\tau_{\text{opt}}$ can be
derived approximately by minimizing $\delta n$ in Eq.
(\ref{delta_n}), which is then given by \cite{Dutta}
\begin{eqnarray}
\tau_{\text{opt}} &\propto & r^{-1/(\beta+1)}=r^{\alpha}.
\end{eqnarray}
This relation is verified to be applicable for all the three
quench protocols in the parameter regions we considered, as
illustrated in Fig. \ref{OptimalTime}. In the original KZM
scenario,  only the control fields without noises accounts for the
production of defect and long quench time $\tau_j$ prevents
defects formation. However, we have shown that the noises in the
control fields, which is a more practical situation in realistic
experiments, can also induce defects and it is intuitively to use
shorter quench time as optimum to suppress the defect production
when noise strength $W$ gets larger.

\section{Quantum simulation of the anti-KZ behavior in two-level systems}

\begin{figure}[tbph]
\centering
\includegraphics[width=7.5cm]{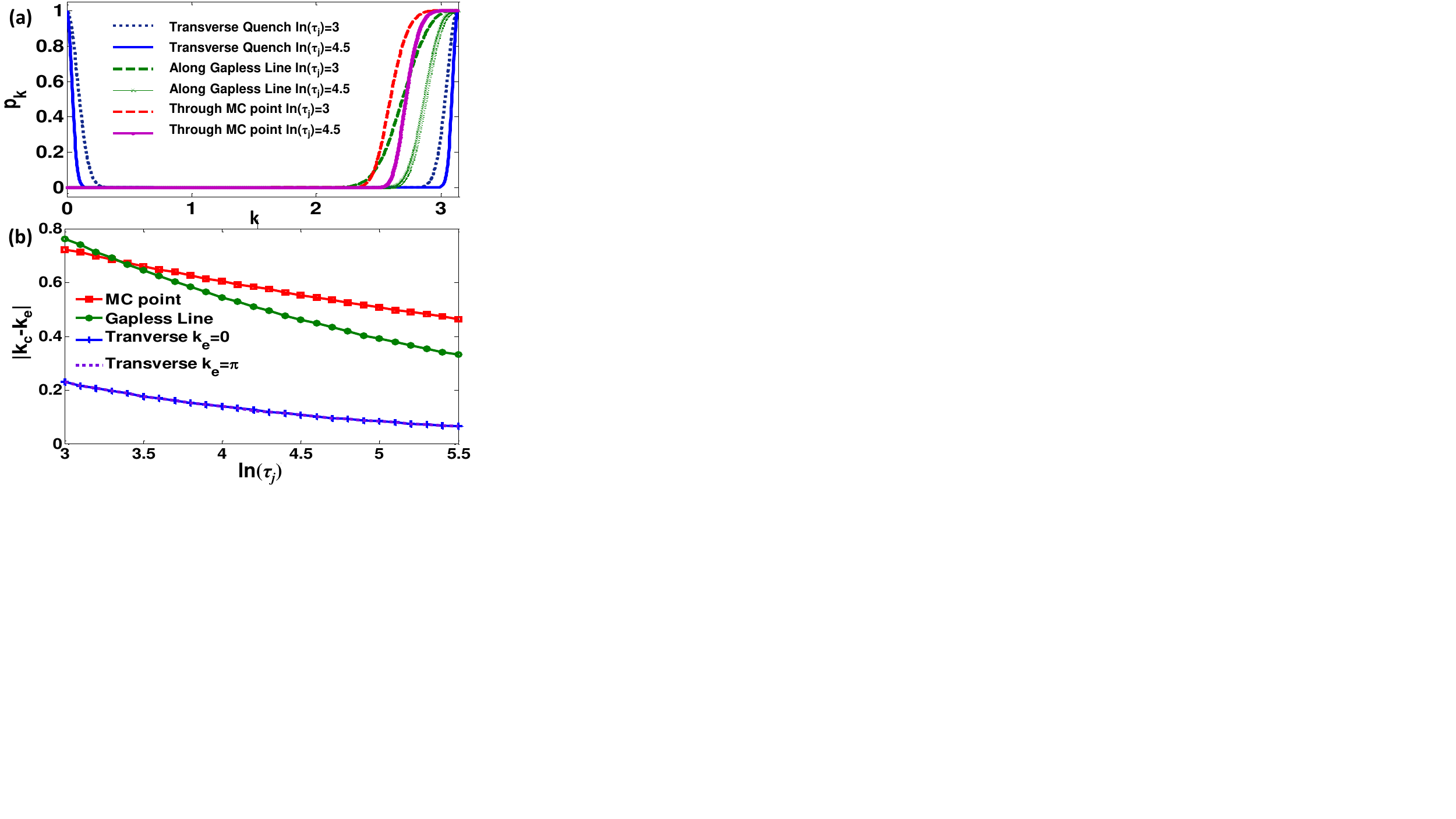}
\caption{(Color online) (a) The excitation probability $p_k$ as a
function of $k$. In the transverse quench ($j=1$), the regions
with major contribution are near the points $k_e=0,\pi$, while
near $k_e=\pi$  in the quench through the multicritical point
($j=2$) and along gapless line ($j=3$). (b) The relative length of
the suggested simulation regions (see the text) $|k_c-k_e|$ as a
function of $\ln(\tau_{j})$, where the cutoff $k_c$ is determined
by the excitation probability $p_k(k)>0.03$.}\label{Pk}
\end{figure}

In this section, we propose to test the predicted anti-KZ behavior
by quantum simulation of the quench dynamics with LZT in two-level
systems. We first present a method to transform a generic
two-level Hamiltonian with time linearly dependent term in the
diagonal term into a standard LZT form. In a two-level system, the
Schr\"{o}dinger equation for the state vector
$|\psi(t)\rangle=u_1(t)|e\rangle+u_2(t)|g\rangle$ with $|e\rangle$
and $|g\rangle$ as the diabatic basis can be written as
\begin{eqnarray}\label{TLS-origin}
i\frac{d}{dt}\left(
\begin{array}{c}
    u_1 \\
    u_2 \\
  \end{array}
\right)
=-2\left(
   \begin{array}{cc}
     vt+C & \Delta \\
     \Delta & -vt-C \\
   \end{array}
 \right)
 \left(
\begin{array}{c}
    u_1 \\
    u_2 \\
  \end{array}
\right).
\end{eqnarray}
Here we assume the parameters $v,\Delta$ and $C$ are real constants. We further use the substitution
\begin{eqnarray}
\nonumber   v_{LZ} &=& v/(2\Delta)^2 \\
  t_{LZ} &=& 4\Delta(t+C/v)
\end{eqnarray}
into the origin two-level system Hamiltonian (\ref{TLS-origin}), which can then be transformed into
the standard LZT form,
\begin{eqnarray}\label{TLS-LZT}
i\frac{d}{dt_{LZ}}\left(
\begin{array}{c}
    u_1 \\
    u_2 \\
  \end{array}
\right)
=-\frac{1}{2}\left(
   \begin{array}{cc}
     v_{LZ}t_{LZ} & 1 \\
     1 & -v_{LZ}t_{LZ} \\
   \end{array}
 \right)
 \left(
\begin{array}{c}
    u_1 \\
    u_2 \\
  \end{array}
\right).
\end{eqnarray}
The probability in the excited state at the end of the driving
is approximately given by the Landau-Zener formula
$P_{LZ}=\exp(-\pi/2v_{LZ})$ \cite{Landau,Zener,Shevchenko}.

On the other hand, KZM can be used to study the dynamics in the
quantum phase transition driven across the quantum critical point.
The essence of KZM is the adiabatic impulse approximation
\cite{LZKZM1}, where the quench process is divided into three
stages:  adiabatic far away from the critical point, frozen state
in the vicinity of the point when $[-\hat{t},\hat{t}]$ with
$\hat{t}$ denoting the freeze-out time scale, and the restart of
adiabatic process. For convention, we define the relaxation time
as $\tau_{\text{rel}}=g^{-1}$, where $g$ is the energy gap between
the ground state and the first excited state, and $\hat{t}$ is
estimated with $\tau_{\text{rel}}=\alpha\hat{t}$ and
$\alpha=\mathcal{O}(1)$. Let $\alpha=1$, the impulse region is
given by $(-v_{LZ}^{-1/2},v_{LZ}^{-1/2})$ \cite{Isingexact1}. When the
starting and ending points of $t_{LZ}$ in the Landau-Zener
Hamiltonian is outside the impulse region, it can be regarded that
there is a complete LZT in this two-level system. The similarity
between LZT and KZM was firstly point out in \cite{LZKZM1}, one of
the most prominent features is that when the system approaches the
critical point, the inverse of the energy gap tends to infinity in
KZM and the counterpart in LZT also increases.

The defect density $n_{W}$ can also be estimated by the integral
of the transition probability $P_{LZ}\approx p_k$  over the
pseudo-momentum space \cite{LZKZMexp_QD,LZKZMexp_SC,LZKZMexp_Ion}
\begin{equation}
n_{W}\approx\frac{1}{\pi}\int_{0}^{\pi}P_{LZ}(k)dk\label{Pkint},
\end{equation}
which can be measured in two-level systems by means of quantum
simulation of the quench dynamics with well-designed Landau-Zener
crossings,  similar as the experiments for the Ising chain without
noises \cite{LZKZMexp_QD,LZKZMexp_SC,LZKZMexp_Ion}. For the three
quench protocols in the noisy XY chain, the parameters in
(\ref{TLS-origin}) for a two-level system correspond to the
counterparts in the Hamiltonian of the noise-free quench
$\mathcal{H}_{j}^{(0)}(k) (j=1,2,3$ for the three protocols). In
addition, the noise part $\eta(t)V_j$ correspond to stochastic
fluctuations of the control fields $V_j$, which can be realized by
inducing the $\hat{\sigma}_z$ or (and) $\hat{\sigma}_x$
fluctuations with tunable strength $W$ into the two-level systems.

To simulate the transverse quench ($j=1$) in transverse field XY
chain with many independent LZT in pseudo-momentum space,  one can
use the mapping $ v \rightarrow v_h$, $\Delta \rightarrow
(J_x-J_y)\sin k$ and $C \rightarrow (J_x+J_y)\cos k$. This
indicates the substitution in this quench protocol
\begin{eqnarray}
\nonumber   v_{LZ} &=& v_h/(2J\gamma\sin k)^2 \\
            t_{LZ} &=& 4J\gamma\sin k(t+J\cos k/v_h),
\end{eqnarray}
which can transform $\mathcal{H}_{1}(k)$ into the standard LZT
form. For the second quench protocol ($j=2$) of anisotropic quench
through the multicritical point, following the mapping of the
parameters similar as those in the first case, one can obtain the
corresponding substitution
\begin{eqnarray}
\nonumber v_{LZ}&=&v_x/[2(J_y\sin 2k+h\sin k)]^2 \\
\nonumber t_{LZ}&=&4(J_y\sin2k+h\sin k) \\
      && [t+(J_y\cos 2k+h\cos k)/v_x].
\end{eqnarray}
Similarly, for the third quench protocol ($j=3$) along the gapless
line, the mapping of the Hamiltonian in pseudo-momentum space
gives the corresponding substitution
\begin{eqnarray}
\nonumber v_{LZ}&=&v_{\gamma}\sin k/[2(\cos k+1)]^2 \\
          t_{LZ}&=&-4(\cos k+1)t.
\end{eqnarray}

To reduce the number of implementing LZT  in the estimation of the
predicted $n_{W}$, we consider the distribution of the excitation
probability $p_k$ as a function of $k$. The results for the three
protocols with typical quench time $\tau_j$ are plotted in Fig.
\ref{Pk}(a). One can find that only those $k$ modes in the regions
near the points $k_e=0$ and (or) $k_e=\pi$ contribute the major to
the excitation formation. Therefore in experiments, one can just
implement some LZT of the $k$ modes in these regions to extract
the simulated defect density. For practical implementation, one
may define a cutoff pseudo-momentum $k_c$ in quantum simulation,
which separates the $k$ axis into two or three (for transverse
quench) parts determined by the excitation probability
$p_k(k)>0.03$ here (other small values are also applicable).
Figure \ref{Pk}(b) depicts $|k_c-k_e|$ as a function of the quench
time in the three protocols, which approximately gives the length
of the required simulation regions measured from the point $k_e$.
The results also show that the region length monotonously
decreases as increasing $\tau_j$.

\section{conclusions}

In summary, we have studied the quench dynamics of a
transverse-field XY chain driven across quantum critical points by
noisy control fields and demonstrated that the defect productions
for three quench protocols with different scaling exponents
exhibit the anti-KZ behavior. We have also shown that
the optimal quench time to minimize defects scales as a universal
power law of the noise strength for all the three cases. Moreover, by
using quantum simulation of the quench dynamics in the spin system
with well-designed Landau-Zener crossings in pseudo-momentum
space, we have proposed an experimentally feasible scheme to test
the predicted anti-KZ behavior.

\acknowledgements{This work was supported by the NKRDP of China
(Grant No. 2016YFA0301803), the NSFC (Grants No.
11604103, No. 11474153, and No. 91636218), the NSF of Guangdong Province (Grant
No. 2016A030313436), and the Startup Foundation of SCNU.}


\begin{thebibliography}{99}

\bibitem{Kibble1976} T. W. B. Kibble, Topology of cosmic domains and strings, J. Phys. A \textbf{9}, 1387 (1976).

\bibitem{Zurek1985} W. H. Zurek, Cosmological experiments in superfluid helium, Nature \textbf{317}, 505 (1985).

\bibitem{Zurek1996} W. H. Zurek, Cosmological experiments in condensed matter systems, Phys. Rep. \textbf{276}, 177 (1996).

\bibitem{Navon2015} N. Navon, A.L. Gaunt, R.P. Smith, and Z. Hadzibabic, Critical dynamics of spontaneous symmetry breaking in a homogeneous Bose gas, Science \textbf{347}, 167 (2015) .

\bibitem{Ulm2013} S. Ulm, J. Ro$\beta$nagel, G. Jacob, C. Deg\"{u}nther, S. T. Dawkins,
U. G. Poschinger, R. Nigmatullin, A. Retzker, M. B. Plenio, F.
Schmidt-Kaler, and K. Singer, Observation of the Kibble-Zurek
scaling law for defect formation in ion crystals, Nat. Commun.
\textbf{4}, 2290 (2013).

\bibitem{Pyka2013} K. Pyka, J. Keller, H. L. Partner, R. Nigmatullin, T. Burgermeister, D. M. Meier, K. Kuhlmann, A Retzker, and M. B. Plenio, Topological defect formation and spontaneous symmetry breaking in ion Coulomb crystals, Nat. Commun. \textbf{4}, 2291 (2013).

\bibitem{Monaco2002} R. Monaco, J. Mygind, and R. J. Rivers, Zurek-Kibble Domain Structures: The Dynamics of Spontaneous Vortex Formation in Annular Josephson Tunnel Junctions,Phys. Rev. Lett. \textbf{89}, 080603 (2002).

\bibitem{Dziarmaga2010} J. Dziarmaga, Dynamics of a quantum phase transition and relaxation to a steady state, Adv. Phys. {\bf59}, 1063 (2010).

\bibitem{Isingexact1} J. Dziarmaga, Dynamics of a quantum phase transition: Exact solution of the quantum Ising model, Phys. Rev. Lett.
\textbf{95}, 245701(2005).

\bibitem{Isingexact2} W. H. Zurek, U. Dorner, and P. Zoller, Dynamics of a quantum phase transition, Phys. Rev. Lett. {\bf95}, 105701 (2005).

\bibitem{Dziarmaga2006} J. Dziarmaga, Dynamics of a quantum phase transition in the random Ising model: logarithmic dependence of the defect density on the transition rate, Phys. Rev. B \textbf{74}, 064416 (2006).

\bibitem{Caneva2007} T. Caneva, R. Fazio, and G. E. Santoro, Adiabatic quantum dynamics of a random Ising chain across its quantum critical point, Phys. Rev. B \textbf{76}, 144427 (2007).


\bibitem{Xy1} V. Mukherjee, U. Divakaran, A. Dutta, and D. Sen, Quenching dynamics of a quantum
XY spin-$1/2$ chain in a transverse field, Phys. Rev. B \textbf{76}, 174303 (2007).

\bibitem{XyMC} U. Divakaran, V. Mukherjee, A. Dutta, and D. Sen, Defect production due to quenching through a multicritical point, J. Stat. Mech. P02007 (2009).

\bibitem{Xy2} U. Divakaran, A. Dutta, and D. Sen, Quenching along a gapless line: A different exponent for defect density, Phys. Rev. B \textbf{78}, 144301 (2008).

\bibitem{Deng2009} S. Deng, G. Ortiz, and L. Viola, Dynamical non-ergodic scaling in continuous finite-order quantum phase transitions, Eurphys. Lett. \textbf{84}, 67008 (2009).

\bibitem{Sabbatini} J. Sabbatini, W. H. Zurek, and M. J. Davis, Phase Separation and Pattern Formation in a Binary Bose-Einstein Condensate, Phys. Rev. Lett. \textbf{107}, 230402 (2011).


\bibitem{Kolodrubetz} M. Kolodrubetz, B. K. Clark, and D. A. Huse, Nonequilibrium Dynamic Critical Scaling of the Quantum Ising Chain, Phys. Rev. Lett. \textbf{109}, 015701 (2012).


\bibitem{Caneva2008} T. Caneva, R. Fazio, and G. E. Santoro, Adiabatic quantum dynamics of the Lipkin-Meshkov-Glick model,
Phys. Rev. B \textbf{78}, 104426(2008).

\bibitem{Acevedo2014} O. L. Acevedo, L. Quiroga, F. J. Rodr¨ªguez, and N. F. Johnson, New dynamical scaling universality for quantum networks across adiabatic quantum phase transitions, Phys. Rev. Lett. \textbf{112}, 030403 (2014).

\bibitem{QPT} S. Sachdev, Quantum Phase Transitions (Cambridge University Press, Cambridge, England, 1999).

\bibitem{Chen2011} D. Chen, M. White, C. Borries, and B. DeMarco, Quantum Quench of an Atomic Mott Insulator, Phys. Rev. Lett. {\bf106}, 235304 (2011).

\bibitem{Braun} S. Braun, M. Friesdorf, S. S. Hodgman, M. Schreiber, J. P. Ronzheimer, A. Riera, M. del Rey, I. Bloch, J. Eisert, and U.
Schneider, Emergence of coherence and the dynamics of quantum phase transitions, Proc. Natl. Acad. Sci. {\bf112}, 3641 (2015).

\bibitem{Anquez} M. Anquez, B. A. Robbins, H.M Bharath, M. Boguslawski, T. M. Hoang, and M. S. Chapman, Quantum Kibble-Zurek Mechanism in a Spin-1 Bose-Einstein Condensate, Phys. Rev. Lett. {\bf116}, 155301 (2016).

\bibitem{Chin} L. W. Clark, L. Feng, and C. Chin, Universal space-time scaling symmetry in the dynamics of bosons across a quantum phase transition, Science {\bf354}, 606 (2016).

\bibitem{Landau} L. D. Landau, On the theory of transfer of energy at collisions II. Physik. Z. Sowjet. {\bf2}, 46 (1932).

\bibitem{Zener} C. Zener, Non-adiabatic crossing of energy levels. Proc. R. Soc. London, Ser. A {\bf137}, 696 (1932).

\bibitem{Shevchenko} S. N. Shevchenko, S. Ashhab, and F. Nori, Landau-Zener-St¨¹ckelberg interferometry. Phys. Rep. {\bf492}, 1 (2010).

\bibitem{Oliver} W. D. Oliver, Y. Yu, J. C. Lee, K. K. Berggren, L. S. Levitov, and T. P. Orlando, Mach-Zehnder interferometry in a strongly driven superconducting qubit. Science {\bf310}, 1653 (2005).

\bibitem{Sillanpaa} M. Sillanpaa, T. Lehtinen, A. Paila, Y. Makhlin, and P. Hakonen, Continuous-time monitoring of Landau-Zener interference in a
cooper-pair box. Phys. Rev. Lett. {\bf96}, 187002 (2006).

\bibitem{Tan} X. Tan, D.-W. Zhang, Z. Zhang, Y. Yu, S. Han, and S.-L. Zhu, Demonstration of Geometric Landau-Zener Interferometry in a Superconducting Qubit. Phys. Rev. Lett. {\bf112}, 027001 (2014).

\bibitem{Petta} J. R. Petta, H. Lu, and A. C. Gossard, A Coherent Beam Splitter for Electronic Spin States, Science {\bf327}, 669 (2010).

\bibitem{Betthausen} C. Betthausen, T. Dollinger, H. Saarikoski, V. Kolkovsky, G. Karczewski, T. Wojtowicz, K. Richter, and D. Weiss, Spin-transistor action via tunable Landau-Zener transitions, Science {\bf337}, 324 (2012).

\bibitem{Cao} G. Cao, H.-O. Li, T. Tu, L. Wang, C. Zhou, M. Xiao, G.-C. Guo, and H.-W. Jiang, Ultrafast universal quantum control of a quantum-dot charge qubit using Landau-Zener-St\"{u}ckelberg interference, Nat. Commun. {\bf4}, 1401 (2013).

\bibitem{Tarruell} L. Tarruell, D. Greif, T. Uehlinger, G. Jotzu, and T. Esslinger, Creating, moving and merging Dirac points with a Fermi gas in a
tunable honeycomb lattice. Nature {\bf483}, 302 (2012).

\bibitem{Salger} T. Salger, C. Geckeler, S. Kling, and M. Weitz, Atomic Landau-Zener tunneling in Fourier-synthesized optical lattices. Phys. Rev. Lett.
{\bf99}, 190405 (2007).

\bibitem{Chen} Y.-A. Chen, S. D. Huber, S. Trotzky, I. Bloch, and E. Altman, Many-body Landau-Zener dynamics in coupled one-dimensional Bose
liquids. Nat. Phys. {\bf7}, 61 (2011).


\bibitem{LZKZM1} B. Damski, The simplest quantum model supporting the Kibble-Zurek mechanism of topological defect production: Landau-Zener transitions from a new perspective, Phys. Rev. Lett. \textbf{95}, 035701 (2005).

\bibitem{LZKZM2} B. Damski and W. H. Zurek, Adiabatic-impulse approximation for avoided level crossings: From phase-transition dynamics to
Landau-Zener evolutions and back again. Phys. Rev. A {\bf73}, 063405 (2006).


\bibitem{Xu} X.-Y. Xu, Y.-J. Han, K. Sun, J.-S. Xu, J.-S. Tang, C.-F. Li, and G.-C. Guo, Quantum simulation of Landau-Zener model dynamics supporting the Kibble-Zurek mechanism. Phys. Rev. Lett. {\bf112}, 035701 (2014).

\bibitem{LZKZMexp_QD} L. Wang, C. Zhou, T. Tu, H.-W. Jiang, G.-P. Guo, and G.-C. Guo, Quantum simulation of the Kibble-Zurek mechanism using a semiconductor electron charge qubit, Phys. Rev. A. \textbf{89}, 022337 (2014).

\bibitem{LZKZMexp_SC} M. Gong, X. Wen, G. Sun, D.-W. Zhang, D. Lan, Y. Zhou, Y. Fan, Y. Liu, X. Tan, H. Yu, Y. Yu, S.-L. Zhu, S. Han, and P. Wu, Simulating the Kibble-Zurek mechanism of the Ising model with a superconducting qubit system, Sci. Rep. \textbf{6}, 22667 (2016).

\bibitem{LZKZMexp_Ion} J.-M. Cui, Y.-F. Huang, Z. Wang, D.-Y. Cao, J. Wang, W.-M. Lv, L. Luo, A. del Campo, Y.-J. Han, C.-F. Li, and G.-C. Guo, Experimental Trapped-ion Quantum Simulation of the Kibble-Zurek dynamics in momentum space, Sci Rep. \textbf{6}, 33381 (2016).

\bibitem{Griffin} S. M. Griffin, M. Lilienblum, K. T. Delaney, Y. Kumagai, M. Fiebig, and N. A. Spaldin, Scaling behavior and beyond equilibrium in the hexagonal manganites, Phys. Rev. X \textbf{2}, 041022(2012).

\bibitem{Novikov}E. A. Novikov, Functionals and the random-force method in turbulence theory, JETP, \textbf{20}, 1290 (1965).

\bibitem{Santoro} D. Patan\`{e}, A. Silva, L. Amico, R. Fazio, and G. E. Santoro, Adiabatic Dynamics in Open Quantum Critical Many-Body Systems, Phys. Rev. Lett. \textbf{101}, 175701 (2008).

\bibitem{Nalbach} P. Nalbach, S. Vishveshwara, and A. A. Clerk, Quantum Kibble-Zurek physics in the presence of spatially correlated dissipation, Phys. Rev. B \textbf{92}, 014306 (2015).

\bibitem{Dutta} A. Dutta, A. Rahmani, and A. del Campo, Anti-Kibble-Zurek Behavior in Crossing the Quantum Critical Point of a Thermally Isolated System Driven by a Noisy Control Field, Phys. Rev. Lett. \textbf{117}, 080402 (2016).

\bibitem{Armin2015} A. Rahmani, Dynamics of noisy quantum systems in the Heisenberg picture: Application to the stability of fractional charge, Phys. Rev. A \textbf{92}, 042110 (2015).

\bibitem{Kim} K. Kim, M. S. Chang, S. Korenblit, R. Islam, E. E. Edwards, J. K.
Freericks, G. D. Lin, L. M. Duan, and C. Monroe, Quantum
simulation of frustrated Ising spins with trapped ions, Nature
{\bf465}, 590 (2010).

\bibitem{Bohnet} J. G. Bohnet, B. C. Sawyer, J. W. Britton, M. L. Wall, A. M. Rey, M. Foss-Feig, and J. J. Bollinger, Quantum spin dynamics and entanglement generation with hundreds of trapped ions, Science \textbf{352}, 1297 (2016).

\bibitem{Simon} J. Simon, W. S. Bakr, R. Ma, M. E. Tai, P. M. Preiss, and M. Greiner, Quantum simulation of antiferromagnetic spin chains in an optical lattice, Nature {\bf472}, 307 (2011).

\bibitem{Lieb1961} E. Lieb, T. Schultz, and D. Mattis, Two soluble models of an antiferromagnetic chain,
Ann. Phys. (N.Y.) \textbf{16}, 407 (1961).

\bibitem{Bunder1999} J. E. Bunder and Ross H. McKenzie, Effect of disorder on quantum phase transitions in anisotropic XY spin chains in a transverse field, Phys. Rev. B \textbf{60}, 344 (1999)



\end{thebibliography}
\end{document}